\documentclass{aa}
\usepackage{epsfig}
\usepackage{natbib}
\usepackage{graphicx}
\bibpunct{(}{)}{;}{a}{}{,}

\usepackage{color}

\usepackage{graphics}
\usepackage{amssymb,amsmath}
\usepackage{amsmath}
\usepackage{psfrag}
\usepackage{graphicx}
\newcommand{\ve}[1]{\mathbf{q1}}

\newcommand{\f}{\frac}

\newcommand{\be}{\begin{equation}}      
\newcommand{\ee}{\end{equation}}      
      
\newcommand{\bef}{\begin{figure}}      
\newcommand{\eef}{\end{figure}}      
\newcommand{\bea}{\begin{eqnarray}}    
\newcommand{\eea}{\end{eqnarray}}

\newcommand{\av}[1]{\ensuremath{\left\langle q1 \right\rangle}}

\newcommand{\tve}[1]{\tilde{\boldsymbol{q1}}}

\def\bse{\begin{subequations}}
\def\ese{\end{subequations}}

\def\lsim{\raise 0.4ex\hbox{$<$}\kern -0.8em\lower 0.62ex\hbox{$\sim$}} 
\def\gsim{\raise 0.4ex\hbox{$>$}\kern -0.7em\lower 0.62ex\hbox{$\sim$}}

\def\f0N{f_0^{(N)}}
\def\bec{\begin{center}}
\def\eec{\end{center}}

\begin{document}

\title{Formation of satellites from {from cold collapse} }
  
\titlerunning{Formation of satellites {from cold collapse} }
  
\authorrunning{Benhaiem \& Sylos Labini}  
  
  \author {David Benhaiem \inst{1,2} and  Francesco  Sylos Labini \inst{3,1,4} }

        \institute{Istituto dei Sistemi Complessi, Consiglio Nazionale
          delle Ricerche, Via dei Taurini 19, 00185 Roma, Italia \and
          Laboratoire de physique nucl\`eaire et de hautes \'energies,
          University Pierre et Marie Curie, 4 Place Jussieu, 75005
          Paris, France Paris, France \and Museo Storico della Fisica
          e Centro Studi e Ricerche Enrico Fermi, Via Panisperna 89 A,
          Compendio del Viminale, 00184 Rome, Italy \and Istituto
          Nazionale Fisica Nucleare, Unit\`a Roma 1, Dipartimento di
          Fisica, Universit\'a di Roma ``Sapienza'', Piazzale Aldo
          Moro 2, 00185 Roma, Italia }

\date{Received / Accepted}

\abstract{ We study the collapse of an isolated, initially cold,
  irregular (but almost spherical) and (slightly) inhomogeneous cloud
  of self-gravitating particles.{The cloud is driven towards a
    virialized quasi-equilibrium state by a fast relaxation mechanism,
    occurring in a typical time $\tau_c$, whose signature is a large
    change in the particle energy distribution.}
Post-collapse particles are divided into two main species: bound and
free, {the latter} being  ejected from the system.
Because of the initial system's anisotropy, the time varying gravitational
field breaks spherical symmetry so that the ejected mass can carry
away angular momentum and the bound system 
can gain a non-zero angular
momentum.
In addition, while strongly bound particles form a compact core,
weakly bound ones may form, in a time scale {of the order of  $\tau_c$},
several satellite sub-structures.
These satellites have a finite lifetime that can be {longer than} $\tau_c$ 
and generally form a {flattened distribution}.
Their origin and their abundance are related to the {amplitude and nature of } 
initial density
fluctuations and to the initial cloud deviations from spherical
symmetry, which are {both} amplified during the collapse phase.
Satellites show a time dependent virial ratio that can be different
from the equilibrium value $b\approx -1$: although they are bound to
the  {main virialized} object, they are not necessarily virially relaxed.
}
\maketitle


\section{Introduction}

The characterization of the collapse of an isolated cloud of
self-gravitating particles represents an important theoretical problem
{that can be possibly relevant for} the comprehension of the
formation astrophysical relaxed objects, as for instance globular
clusters, galaxies etc.  It is known, since the first numerical
simulations, that when an isolated self-gravitating particles system
is initially cold, it collapses {violently and then it} relaxes to
produce a virialized quasi-equilibrium state {in a relatively
  short time scale} \citep{henon_1964}. While the collapse of an
uniform spherical cloud was {the case} mostly studied in the
literature (see e.g., \cite{aarseth_etal_1988,boily_etal_2002} and
references therein), many authors have focused on spherical models
with non-trivial density profiles and/or with significant non-zero
velocities {(see, e.g.,
  \cite{vanalbada_1982,mcglynn_1984,villumsen_1984,aguilar+merritt_1990,
    theis+spurzem_1999,merrall+henriksen_2003,roy+perez_2004,boily+athanassoula_2006}
  and references therein)}.

A crucial parameter determining the cloud evolution after the collapse
is { the initial virial ratio,} $b(0) = {2K}/{W}$ i.e., the ratio
between twice the kinetic energy $K$ and the gravitational potential
energy $W$.  When the cloud is cold enough (i.e., $b(0) \rightarrow
0$) it can change its size by a relevant factor, passing hence
{through a phase of fast relaxation whose} signature is
represented by a large change of the particle energy distribution
$P(e_p)$: {while initially all particles had negative energy,
  after the collapse two different species of particles arise: bound
  particles and free particles
  \citep{david+theuns_1989,theuns+david_1990,ejection_mjbmfsl,syloslabini_2012}.}
Bound particles, with energy $e_p \ll 0$, form the core of the
virialized object; weakly bound particles, with energy $e_p \le 0$,
form the large radii tail of the virialized {object;} free
particles, with energy $e_p >0$, can escape from the system.  On the
other hand, when the cloud is initially warm enough (i.e., $b(0)
\rightarrow -1$) it slightly rearranges its phase-space distribution
in order to reach a quasi stationary state without ejecting mass and
energy {(see, e.g., \citet{syloslabini_2012} and references
  therein)}.

{The fast relaxation} mechanism is associated with a number of
interesting features: {beyond the ejection of mass and energy from
  the system one may observe the formation} of a virialized
quasi-stationary state with universal density and velocity profiles.
{This occurs because of} the strong gravitational field generated
during the collapse washes out the dependence from initial conditions
\citep{syloslabini_2013}.
{Several authors (e.g.,
  \cite{merritt+aguilar_1985,aguilar+merritt_1990,theis+spurzem_1999,
    boily+athanassoula_2006,barnes_etal_2009,syloslabini_etal_2015})
  have studied the breaking of the initial spherical symmetry in
  consequence of the collapse}.  {Recently,
  \cite{worrakitpoonpon_2014,Benhaiem_etal_2016} noticed that the
  virialized state may have a non-zero angular momentum}. Indeed,
together with mass and energy, ejected particles can carry away
angular momentum if spherical symmetry is broken during the
gravitational collapse.

Beyond the case of a spherical cloud of self-gravitating particles,
{we have studied the dynamics of the cold collapse} by using initially
cold and {slightly} ellipsoidal
clouds~\citep{Benhaiem+SylosLabini_2015}. This system was chosen
because it represents a relatively simple configuration for
determining the effects of the deviations from spherical symmetry
{during the cold collapse}. While the main characteristics of the
virialized object are {very similar to the spherical cloud case}, {we}
found that ejected particles are characterized by highly asymmetrical
shapes, whose features can be traced in the initial deviations from
spherical symmetry that are amplified during the {cold collapse}
phase. Ejected particles can indeed form {flattened} configurations
even though the initial cloud was very close to spherical.
Furthermore we noted that ellipsoidal clouds {with a sufficiently
  large initial degree of spatial anisotropy} (i.e., the ratio between
the largest and smallest semi-axis smaller than $\approx 0.8$) may
give rise to the formation of satellite sub-structures that are not
ejected from the system and that ultimately fall back on the main
virialized object. {In addition, these sub-structures are found to
  follow the same spatial distribution of the ejected particles and
  thus forming flattened configurations.}

{This observation motivated the present paper where we consider more
  inhomogeneous initial conditions aiming to determine how these might
  favor the formation of satellites sub-structures during a cold
  collapse}.  We {thus} consider hereafter the collapse of isolated,
cold, irregular (but almost spherical), (slightly) inhomogeneous
self-gravitating clouds.  { \cite{vanalbada_1982} performed some
  numerical experiments by using similar initial conditions to ours;
  here {we study in greater details the the morphology of the
    aggregates of bound particles} that are formed after the collapse
  as well as the formation of satellite sub-structures and a number of
  other interesting physical features, thanks to much more powerful
  computational means}.

{The paper is organized as follows.  {In}
  Sect.\ref{simulations} we illustrate the simulations that we have
  performed, giving details on the {generation of the} initial
  conditions, {the numerical codes and algorithm used to
    identified the satellites.}  The main results of our numerical
  experiments are presented in Sect.\ref{results}.  {Finally in
    Sect.\ref{conclusion} we draw our main conclusions.}  }

%%%%%%%%%%%%%%%%%%%%%%%%%%%%%%%%%%%%%%%%%%%%%%%%%%%%%%%%%%%%%%%%%%%%%%%%%%%%%%%%%%

\section{The Simulations} 
\label{simulations}
%%%%%%%%%%%%%%%%%%%%%%%%%%%%%%%%%%%%%%%%%%%%%%%%%%%%%%%%%%%%%%%%%%%%%%%%%%%%%%%%%%

\subsection{Generation of initial conditions}

The weakly inhomogeneous and slightly non-spherical particle
distributions used as initial conditions are generated as follows. We
randomly place $N_c$ points in a sphere of radius $R_0$.  {
Each of these particles is taken to be the center
of a spherical cloud of $N_p$ particles, where for each case
the number $N_p$ is extracted from a uniform distribution
with average $\mu$ and variance $\sigma$. Thus the expected total
number of particles is thus $N = N_c  \times \mu$}. {The $N_p$ particles
are randomly distributed in each sub-cloud that is taken
to be spherical and with radius $l_c$}: this is a free parameter that can
  be adjusted as explained below.  All particles have the same mass
  $m$ and the initial cloud density is approximately
\be
\label{rho0}
\rho_{0} \approx \frac{N m}{4\pi/3 R_0^3} \;, 
\ee
because the initial system is close to spherical as long as $l_c <
R_0$.  Similarly the collapse time\footnote{In our units $\rho_0=1
  gr/cm^3$ and thus $\tau_c=2100$ sec.} of the whole cloud is of the
order of that of a perfectly spherical one with density $\rho_0$,
{i.e.,
$\tau_c \approx \sqrt{ {3 \pi }/({32 G\rho_0} )}  \;. $
%\ee
%
}

{The radius of each sub-cloud $l_c$ is fixed to be larger than the
  average distance between sub-clouds centers}
  \be
  \l_c  \ge \Lambda_c = \left( \frac{4 \pi R_0^3/3}{N_c} \right)^{1/3} \;.
  \ee
{In such a way the different sub-clouds may substantially overlap so that the resulting density distribution does not have large fluctuations  (i.e., there are not empty regions)}.  The collapse time of each
  sub-cloud, if it were isolated, is 
%
%\be
$\tau_p = \sqrt{ {3 \pi }/({32 G \rho_p} )}  \;, $
%\ee
where $\rho_p = 3mN_p/(4\pi l_c^3)$.  % (see Tab.\ref{table_cl}).
When $l_c > \Lambda_c$ we find $\tau_p > \tau_c$, i.e.  the collapse
of the whole cloud occurs before than the collapse of each
sub-cloud. This means that the mean field of the inhomogeneous and
non-spherical cloud drives the dynamics of the {monolithic} collapse
of the whole system.

Otherwise when $l_c < \Lambda_c$ the collapse of the sub-clouds is
faster than that of the whole cloud, i.e. $\tau_p < \tau_c$.
{However,} in this case the initial distribution {is } highly
inhomogeneous: each spherical sub-cloud collapses independently of all
others, { forming a virialized state that may eventually} merge
with the others, {giving rise to a hierarchical bottom-up
  aggregation process}. Here we limit our study to the case of weakly
inhomogeneous systems that give rise {to monolithic} collapses, { and
  thus we fix in all our simulations $l_c \ge \Lambda_c$}.

{In order to} characterize a {generic} structure shape we
define three different linear combinations of the three eigenvalues
$\lambda_i$ (defined as $\lambda_1 \ge \lambda_2 \ge \lambda_3$) of
the inertia tensor {\citep{binney}}: the flatness parameter
\be 
\label{iota} 
\iota = \frac{\lambda_1}{\lambda_3} -1 \;, 
\ee
the triaxiality index
\be 
\label{tau} 
\tau = \frac{\lambda_2-\lambda_3}{\lambda_1 - \lambda_3} 
\ee
and the disk parameter 
\be 
\label{phi} 
\phi = \frac{\lambda_1-\lambda_3}{\lambda_2+ \lambda_3} \;.
\ee
The combination of $\iota, \tau$ and $\phi$ allows one to distinguish
not only between different types of ellipsoids (prolate, oblate and
triaxial) but also between other shapes like bars and disks
{\citep{Benhaiem+SylosLabini_2015}}.  In
{Table}~\ref{table-param} we report their values for some {
  cases, representative of the range of parameters we explore in our
  numerical experiments}.
\begin{table}
\begin{center}
\begin{tabular}{|c|c|c|c|c|}
  \hline
Name           & $\iota$  & $\tau$   & $\phi$                       \\%&   $\Sigma$        \\ 
\hline
Sphere         &   0        &  --    & 0                            \\%&   --             \\ 
Prolate        &   $\iota$  &  1     & $\approx \frac{\iota}{2}$    \\%&   $\frac{3}{2}\iota+1$    \\ 
Oblate         &   $\iota$  &  0     & $\approx  \frac{\iota}{2}$   \\%&   $\frac{3}{2}\iota$      \\
Triaxial       &   $\iota$  &  1/2   & $\frac{2\iota}{4+\iota}$     \\%&   $\frac{6 \iota + \iota^2} {4+\iota} +  \frac{1}{2}$  \\  
Disk           &   1        &  0     & $\frac{1}{2}$                \\%&   $\frac{3}{2}$             \\
Tiny Cylinder  &   $\gg 1$  &  1     & $\approx 1$                  \\%&   $\gg 3$         \\
\hline
\end{tabular}
\end{center}
\caption{Values of the parameters $\iota,\tau,\phi$ for known shapes.}
\label{table-param}
\end{table}
    {Then, in} Tab.\ref{table_cl} we show the details of the
    simulations we have run.

    {We emphasize that the initial
      velocity dispersion was set to zero, i.e. we have considered
      perfectly cold initial conditions. As discussed in
      \cite{syloslabini_2012} a monolithic collapse occurs as long as
      the initial velocity dispersion is small enough, i.e. for
      an initial virial ratio such that} 
      \be
      b = \frac{2W}{K} > -\frac{1}{2} \;,
      \ee
{where $K(W)$ is the initial total kinetic (potential) energy.}
      {Indeed if the initial velocities are larger the
        monolithic collapse is not violent anymore, i.e. the particle
        energy distribution does not change in a relevant way.  As we
        discuss below the generation of satellites is instead related
        to a large particle energy change during the collapse and for
        this reason we consider only cold initial conditions. In
        particular we take the initial virial ration to be $b=0$.}

\begin{table}
\begin{center}
\begin{tabular}{c|c|c|c|c|c|c|c|c|}
\hline 
Name       &  $N_c$      & $n$ &  $l_c/\Lambda_c$ & $\tau_p/\tau_c$     & $N_s$       & $\iota$  & $\tau$   & $\phi$    \\
\hline
C1           &   10      &   36          &  3            & 3                &  3        &  0.2     & 0.7     & 0.1      \\  %10000.1 
C2           &   10      &   36          &  1            & 1                &  2        &  0.5     & 0.9     & 0.2     \\   %10000.2 
%10000.3 clustered
%10000.4 clustered
%10000.5 clustered
%
C3           &   50      &  22          &  4             & 7                &  0        &  0.1      & 0.2      & 0.1        \\   %10000.6 
C4           &   50      &  23          &  2             & 3                &  1        &  0.2      & 0.3      & 0.1        \\   %10000.7 
%10000.8 clustered
C5           &   100     &  28         &  5             & 10                &  1        &  0.6      & 0.3      & 0.03       \\  %10000.9 
C6           &   10      &  330        &  1             & 1                 &  5        &  0.2      & 0.7      & 0.1        \\  %100000.1
C7           &   10      &  330        &  1             & 1                 &  14       &  0.4      & 0.9      & 0.2        \\  %100000.2 
C8           &   3       &  570        &  2             & 1                 &  9        &  0.2      & 0.6      & 0.1        \\  %100000.3 
C9           &   3       &  570        &  1             & 1                 &  8        &  0.7      & 0.7      & 0.3        \\  %100000.4
C10          &   10      &  330        &  5             & 9                 &  5        &  0.04     & 0.8      & 0.02       \\  %100000.5
C11          &   20      &  261        &  12            & 13                &  5        &  0.04     & 0.7      & 0.02       \\  %100000.6 
C12          &   20      &  261        &  12            & 35                &  9        &  0.01     & 0.8      & 0.06       \\  %100000.7 
C13          &   40      &  208        &  4             & 6                 &  2        &  0.2      & 0.5      & 0.1        \\  %100000.8 
C14          &   40      &  208        &  12            & 33                &  3        &  0.03     & 0.6      & 0.01       \\  %100000.9 
C15          &   40      &  208        & 2              & 2                 &  7        &  0.4      & 0.5      & 0.2        \\  %100000.10 
C16          &   5       &  159        & 2              & 2                 &  8        &  0.2      & 0.9      & 0.1        \\  %100000.11  
C17          &   5       &  159        & 6              & 11                &  3        &  0.03     & 0.9      & 0.01       \\  %100000.12  
C18          &   5       &  159        & 1              & 1                 &  6        &  0.7      & 0.9      & 0.3         \\  %100000.13  
C19          &   10      &  180        & 8              & 16                &  6        &  0.2      & 0.3      & 0.1         \\  %100000.14  
C20          &   10      &  181        & 3              & 3                 &  8        &  0.1      & 0.3      & 0.07       \\  %100000.15  
C21          &   10      &  181        & 1              & 1                 &  9        &  0.3      & 0.3      & 0.1        \\  %100000.16  
%C22          &   20      &  181        & 10            & 23                 &  ?       &           &          &              \\  %100000.17
C22          &   20      &  181        & 3              & 4                 &  4        &  0.1      & 0.3      & 0.06       \\  %100000.18  
C23          &   20      &  186        & 2              & 2                 &  3        &  0.3      & 0.6      & 0.2        \\  %100000.19  
\hline 
\end{tabular}
\end{center}
\caption{Properties of inhomogeneous cloud simulations. The columns
  are: $N_c$ number of centers, {$n = N/10^3$ where $N$ is the total number of
  particles}, $l_c/\Lambda_c$ ratio between the size of a
  {sub-cloud} and the average distance between centers,
  $\tau_p/\tau_c$ ratio between the time of collapse of a
  {sub-cloud} and that of the cloud, $N_s$ is the number of
    satellite after the collapse { and $\iota, \tau$, $\phi$ are
      the shape parameters of the initial conditions.}}

%
%Simulations with cluster initial conditions (created with the
%  code {\tt Clusters.f}). for the series 100001.1, ..., we used the
%  code {\tt Clusters2.f} }
\label{table_cl}
\end{table}

%%%%%%%%%%%%%%%%%%%%%%%%%%%%%%%%%%%%%%%%%%%%%%%%%%%%%%%%%%%%%%%%%%%%%%%%%%%%%%%%%%

\subsection{Numerical Details}
\label{sec:numerical} 

In order to run gravitational N-body simulations we have used the
N-body code { {\tt Gadget-2}
  \citep{gadget_paper,Springel_etal_2005} (without the Fourier
  transform part that is unnecessary for an isolated system)}.  All
results presented here are for simulations in which energy is
conserved to within a few tenths of a percent over the time scale
evolved, with maximal deviations at any time of less than half a
{percent.  This accuracy} has been attained using values of the
essential numerical parameters in the { {\tt Gadget-2}} code
$0.025$ for the $\eta$ parameter controlling the time-step, and a
force accuracy parameter of $\alpha_F= 0.001$, that is in the range of
typically used values for this code.  We have also performed extensive
tests of the effect of varying the force smoothing parameter
$\varepsilon$, and we found that we obtain very stable results
provided it is significantly smaller than the minimal size reached by
the whole structure during collapse.  Our simulations are performed
with open boundary conditions and in a non-expanding background. Thus,
beyond energy conservation, we can also monitor linear and angular
momentum conservation to test the accuracy of the numerical
integration.  Simulations are run up to {$\sim 20 \tau_c$}, which
is a {long enough} time range {to study the structures
  emerging from a monolithic collapse, and it requires long
  calculations,} in particular, for simulations with $>10^5$
particles. 
{Indeed, in such a time scale the structure formed from the
  monolithic collapse has sufficient time to relax to a quasi
  stationary state which has (almost) time-independent properties
  \cite{syloslabini_2012}. Then there are still two main sources of
  the further evolution taking place at longer times. On the one hand,
  the bound satellites orbiting around the main structure may cross
  it once or several times up to when they will be destroyed by tidal
  interactions. As mentioned above, such a dynamical process can take
  a very long time scale. On the other hand  two-body collisions  will 
  cause a slow evaporation
  of the virialized object in a  typical time scale of the
  order of $N/\log(N) \tau_c$, i.e.  very much longer than the time
  scale we have explored.}

%%%%%%%%%%%%%%%%%%%%%%%%%%%%%%%%%%%%%%%%%%%%%%%%%%%%%%%%%%%%%%%%%%%%%%%%%%%%%%%%%%

\subsection{Satellites identification} 
\label{subsec:satellites} 

As mentioned above, after the {the cold collapse} phase we detect
in all cases the formation of satellites, i.e. aggregates of particles
that can be self-bound, at least for a finite time, and that are bound
to the main object {that} contains most of the particles.  In
order to identity these sub-structures in the simulations, we have
developed a simple algorithm, {which is inspired by the
  {spherical overdensity} algorithms}, {usually} used in
cosmological simulations to find the so-called ``sub-halos'' {(see
  \citet{onions2012subhaloes} and references therein}).  This
algorithm has been extensively tested both against some artificial
configurations that we used as toy models and against the
gravitational simulations we have run.  In particular, we have checked
that the satellites identified by the algorithm correspond to those
that we visually identify and that these are very robust to the change
of the possible numerical parameters of the code, i.e. they are
consistent with respect to the possible arbitrary choices of the
different numerical values used by the algorithm.

{Our algorithm finds satellite sub-structures in a certain
  snapshot of a given run: as long as satellites are sufficiently far
  away from the largest virialized object, the results are very weakly
  dependent on the time chosen, which in general {do not exceed
    several collapse's characteristic time, i.e. $5\div10$ $\tau_c$.}
  The code} firstly identifies the position of the main object and the
particles that belong to it. To this aim, we compute the potential
energy of each particle and we define the center of the main object as
the particle with the minimum of the gravitational potential
energy. Starting from this point we compute the radial gravitational
potential's profile of the whole distribution.  If there are no
sub-structures, the potential should smoothly decay from the object's
center.

We define the radius of the main object as the radius $r_{mo}$ at
which we observe either: (1) that the absolute value of the radial
potential increases over two consecutive radial bins or, (2) that the
mean density at {$r_{mo}$ is} equal to a certain density
threshold.  For instance we use
%
%\[
$ \rho_{thresh}=0.1 \rho_0,$
%\] 
%
where $\rho_0$ is the initial density of the structure {(see
Eq.\ref{rho0})}. We then tag all the particles within this radius as
  belonging to the main object; they are then stored and put aside
  from the subsequent calculations.
This procedure is {repeated} iteratively: we first remove all the
particles belonging to the main object and then we compute again the
potential of all remaining particles. We then find the next minimum of
the gravitational potential, and we use the same procedure illustrated
above to find the radius of the satellite. We tag the particle
belonging to the first satellite accordingly and remove it from the
next calculation. We proceed in this way until a satellite has a
number of particles below a threshold that we set equal to
$N_{min}=100$: at this point the extraction of the satellites is
stopped.

{{We have tested different values of $\rho_{thresh}$ to find the
    best one for a given simulation. Indeed, a too high density
    threshold artificially reduces the size of the main object, thus
    leading to the false identification of particles as belonging to a
    satellite while they are in reality embedded in the main object
    itself.  On the other hand a too small density threshold would
    include also escapers particles as satellites.  Instead, we tune
    the density threshold to detect satellites with sufficient number
    of particles and dense enough that they can survive during a few
    dynamical times; we have then checked very carefully by visual
    inspection (and by checking that the particles identified as
    belonging to a satellite remains mostly the same at different
    times) that it was indeed the case.}

After the identification of the main object and of its satellites, we
proceed with the analysis to characterize their properties.  First we
compute both the density and the velocity profiles.  Then we estimate
their virial ratio
%
%\be 
$b_s={2K_s}/{W_s} \;,$
%\ee 
%
where $W_s$ is computed by considering only the particles of each
(sub)structure and $K_s$ is computed in the center of mass rest frame.
Further we compute {their angular momenta}  
and we characterize their
shapes by measuring the  parameters $\iota, \tau$ and $\phi$ (see
Eqs.~\ref{iota}-\ref{tau}).

In addition, we compute the equation of the plane identified by the
major and minor semi axis of the main object and then we compute the
distance of the satellites from this plane.  In this way we can easily
determine the angle between the satellite center of mass and such a
plane. {This measurement allows us to understand whether the
  satellites form a flattened distribution.}

%%%%%%%%%%%%%%%%%%%%%%%%%%%%%%%%%%%%%%%%%%%%%%%%%%%%%%%%%%%%%%%%%%%%%%%%%

\subsection{Effects of the initial inhomogeneities on the demographics
of the satellites}

{The relation between the characteristics of the 
  initial spatial configurations, as for the examples the number of 
sub-clouds and the degree of anisotropy, and the number of
satellites formed after the monolithic collapse of the cloud
is not regulated by a few simple macroscopic parameters.
Indeed, we cannot detect a correlation with the number of
satellites detected after the collapse neither with the 
initial number of sub-clouds nor with the the initial values
of the structure parameters $\iota,\tau$ and $\phi$ (see Tab.\ref{table_cl}).}

  {In addition, we have tried to assest the role played by
the initial density inhomogeneities in determining the final
number of satellites formed after the monolithic collapse
by applying the satellite finding algorithm to the initial
conditions. Even in this case we do not detect any corre-
lation as the algorithm that we used for the analysis of
the evolved snapshots does not identify any satellite in
the initial conditions. This is because the code identifies a
satellite if this is a well-defined and isolated object while
the initial conditions we considered are instead sufficiently
uniform and do not contain such objects. As discussed be-
low, the formation of satellites is driven by a non-trivial
  combination of the roles of the initial density 
fuctuations
and of the initial spatial anisotropy and for this reason,
for small deviations from uniformity and spherical sym-
metry, there is not a simple a definite initial characteristic
of the cloud which can be correlated with the number of
post-collapse satellites.}

%%%%%%%%%%%%%%%%%%%%%%%%%%%%%%%%%%%%%%%%%%%%%%%%%%%%%%%%%%%%%%%%%%%%%%%
%%%%%%%%%%%%%%%%%%%%%%%%%%%%%%%%%%%%%%%%%%%%%%%%%%%%%%%%%%%%%%%%%%%%%%%

\section{Results} 
\label{results} 

Let us firstly focus on some global properties of a typical
simulation. As one may see in Fig.\ref{fig1} (upper panel) the total
energy is conserved up to 0.2\%, {with an absolute maximum} at the
time of global collapse of the system, that corresponds to a deviation
of $0.6\%$. This is in line with the fact that the numerical
integration is more difficult when the density is higher: however
collisions do not represent a major issue because {the typical
  time-scale for two-body relaxation is much larger than the duration
  of the collapse phase $\tau_c$.}
The difference between the virial ratio of all particles and that of
all bound particles is due to the ejection of mass from the system,
which is very small in the typical case we consider in this paper (see
the bottom panel of Fig.\ref{fig1}).

\begin{figure}
\vspace{1cm} { \par\centering
  \resizebox*{9cm}{8cm}{\includegraphics*{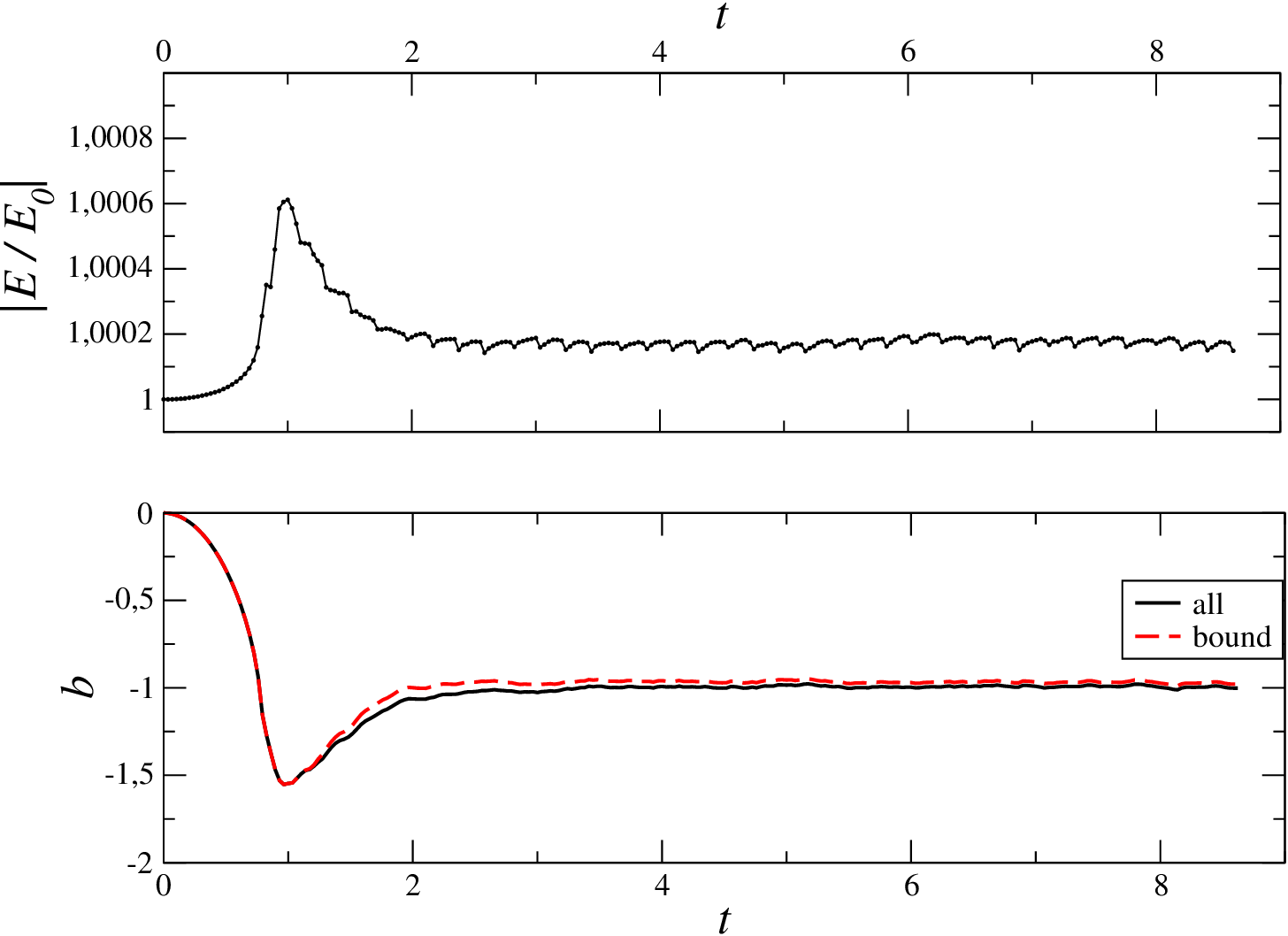}} 
\par\centering }
\caption{Simulation C1. Upper panel: total energy. Bottom panel:
  virial ratio of all particles (solid line) and of bound ones (dashed
  line). }
\label{fig1}
\end{figure}
%%%%%
{We first concentrate our study to the properties of the main
  object, extracted through the procedure described in
  Section.\ref{subsec:satellites}, then we analyze the weakly bound
  particles surrounding the main structure. Finally we investigate the
  properties of the satellites and how they are distributed.}

%%%%%%%%%%%%%%%%%%%%%%%%  MAIN STRUCTURE %%%%%%%%%%%%%%%%%%%%%%%%%%%%%%%%%%%%%%%%%%%
\subsection{Main virialized object} 

The largest virialized object {is in a quasi stationary state} for
$t > \tau_c$ and {it} is almost triaxial because we {find}
$\phi \approx 1/2$ and $\tau \approx 0.2$ being $\iota \approx 0.4$
(see Fig.\ref{fig2}).  {Fig.\ref{cluster_1e4_2} shows the density
  profile that shows a power-law tail decaying as $r^{-4}$ (left
  panel) and the radial velocity dispersion profiles that shows a tail
  decaying as $r^{-1}$ (right panel).  }
\begin{figure}
\vspace{1cm} { \par\centering
  \resizebox*{9cm}{8cm}{\includegraphics*{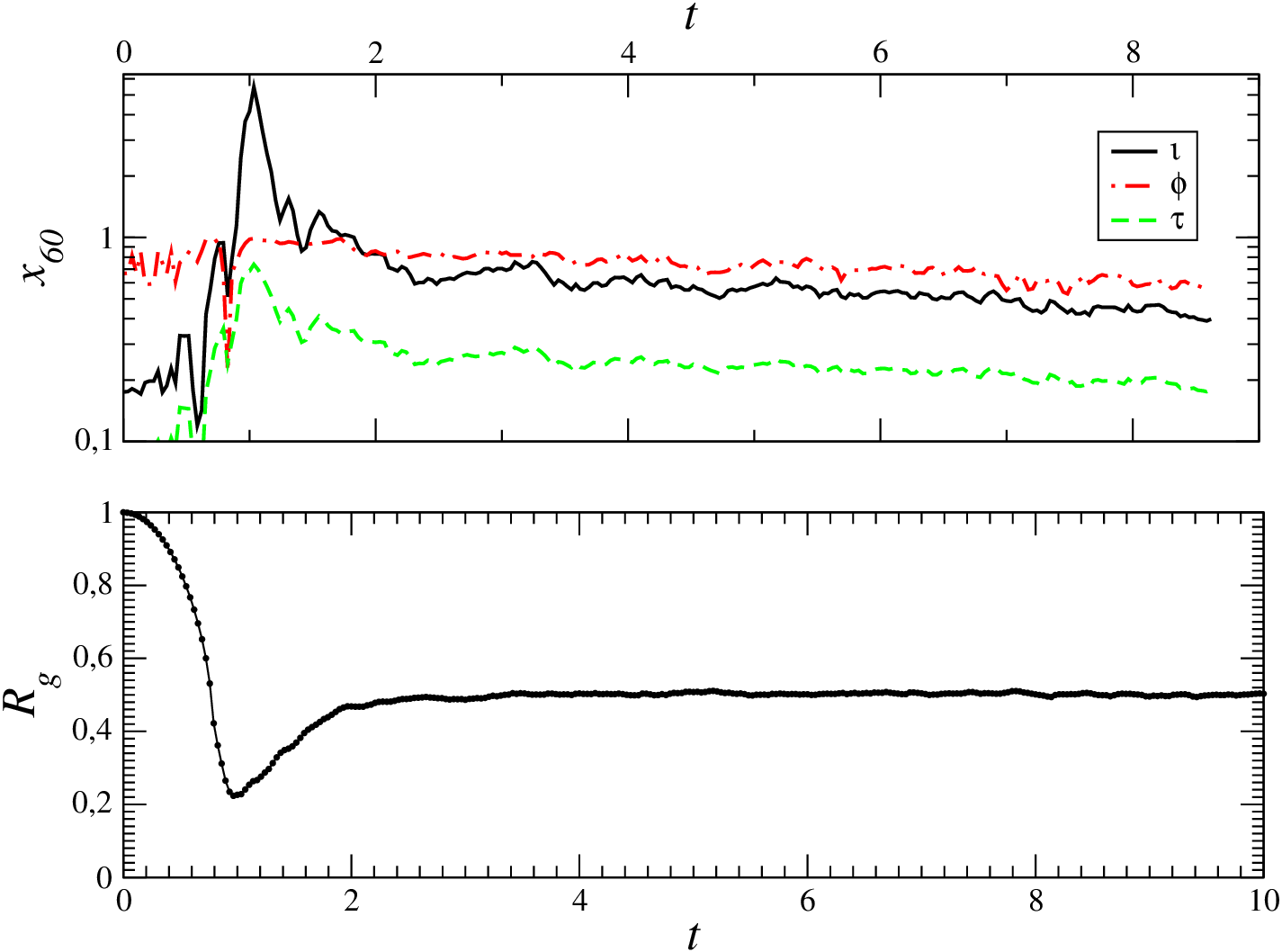}} 
\par\centering }
\caption{Simulation C1. Upper panel: behavior of the shape parameters
  $\iota,\phi,\tau$ for the 60\% more bound particles of the largest
  virialized structure. Bottom panel: Behavior of the gravitational
  radius.}
\label{fig2}
\end{figure}
%%%%%figure7
%
%
\begin{figure}
\vspace{1cm} { \par\centering
  \resizebox*{9cm}{4cm}{\includegraphics*{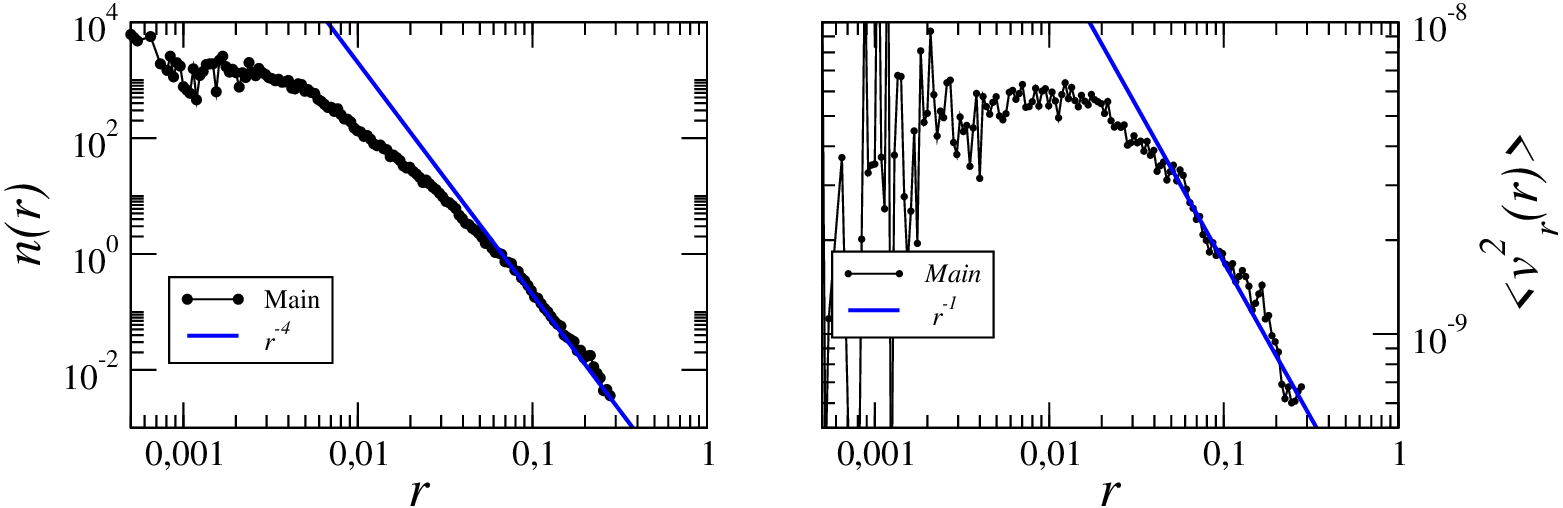}} \par\centering
}
\caption{Simulation C1. Left panel: density profile ({the solid
line is a power-law with an exponent $-4$}). Right panel:
  radial velocity dispersion profile ({the solid
line is a power-law with an exponent $-1$}).}
\label{cluster_1e4_2}
\end{figure}

{During the collapse, all particles move almost radially towards
  the center, up to the arrival time when their velocity is mostly
  directed outwards: particles have different arrival times depending
  on their initial position.}  For inhomogeneous and non-spherical
clouds the spread {of arrival times} is much larger than {in} the
case in which the initial condition was {either} spherical {or}
{slightly} ellipsoidal.  {In the spherical case the spread of
  arrival times can be tuned by changing} the initial density profile
{shape}: {the minimal spread occurs for the case of a uniform
  density profile.}  {On the other hand, when} the initial
{density} profile is power-law, or {when} it has more
complicated behaviors, then the arrival times {spread} can be
%very {much} larger.  {The presence of this spread} implies
that particles {initially in } the {most} external shells
arrive at the center when {the} {particles,} initially in the
internal shells, are already re-expanding: {it is the motion in a
  rapidly varying potential that allows {these} particles to gain
  kinetic energy (see \cite{ejection_mjbmfsl} for a quantitative
  discussion of this phenomenon), causing a large variation of the
  particle energy distribution $P(e_p)$. }  { In particular this
  occurs, if the cloud's minimal gravitational radius (in units of the
  initial one) is
\be
{R}_g(t) = \frac{W(t=0)}{W(t)} \le 0.5 
\ee
(where $W(t)$ is the gravitational potential energy at time $t$ ---
see the bottom panel of Fig.\ref{fig2}).  This criterion was found to
hold also when the initial cloud has a uniform density profile but an
ellipsoidal shape (with $\iota(0) \le 0.25 $ --- see
\cite{Benhaiem+SylosLabini_2015}) and it works well also in the case
of the simulations presented hereafter.}

\subsection{Weakly bounded particles} 

As mentioned above, the collapse of {a} slightly inhomogeneous and
weakly non-spherical cloud presents features {in common with both
  initial spherical and ellipsoidal ones.  In particular, }in the
simulations we have performed, we find that a large fraction of the
mass collapses in a small time interval around $\tau_c$ and then the
more external shells collapse with a large spread of arrival times.
In this situation, because spherical symmetry is  {already broken} in the
initial conditions, not only there is a large variation of
{individual particle energies} but also other interesting features
arise {such} as an asymmetric ejection of mass and the formation
of a {flattened} configuration of weakly bounded particles. In addition
we find that some of these weakly bounded particles aggregate into
satellites sub-structures while others may form time-dependent
configurations like streams, jets and rings.

The flattening of the weakly bound particles distribution is shown by
comparing the behavior of the ratio between the minor to major
eigenvalue at time $t^*> \tau_c$ with that of the initial cloud (see
Fig.\ref{c_over_a}).
{As mentioned above after a few dynamical times after the
  monolithic collapse the majority of the bound particles relax to a
  quasi-stationary state whose properties change over a very long time
  scale that is much longer than what we consider in our numerical
  experiments. Indeed in Figs.\ref{fig1}-\ref{fig2} one may note that
  the macroscopic characteristic of the structure are almost time
  independent. Therefore for most of the particles the ratio between
  the minor to major eigenvalue is time independent. On the other hand,
  a fraction of the bound particles, those with energy close to
  zero, still evolve after the collapse for a time which depends on
  their energy and can be much longer than the time scale of our
  simulations. We have controlled that in the time range $5-20 \tau_c$
  the ratio between the minor to major eigenvalue does not change
  significantly and thus we report the measurement of this ratio at
  $10 \tau_c$.}

\begin{figure}
\vspace{1cm} { \par\centering
\resizebox*{9cm}{8cm}{\includegraphics*{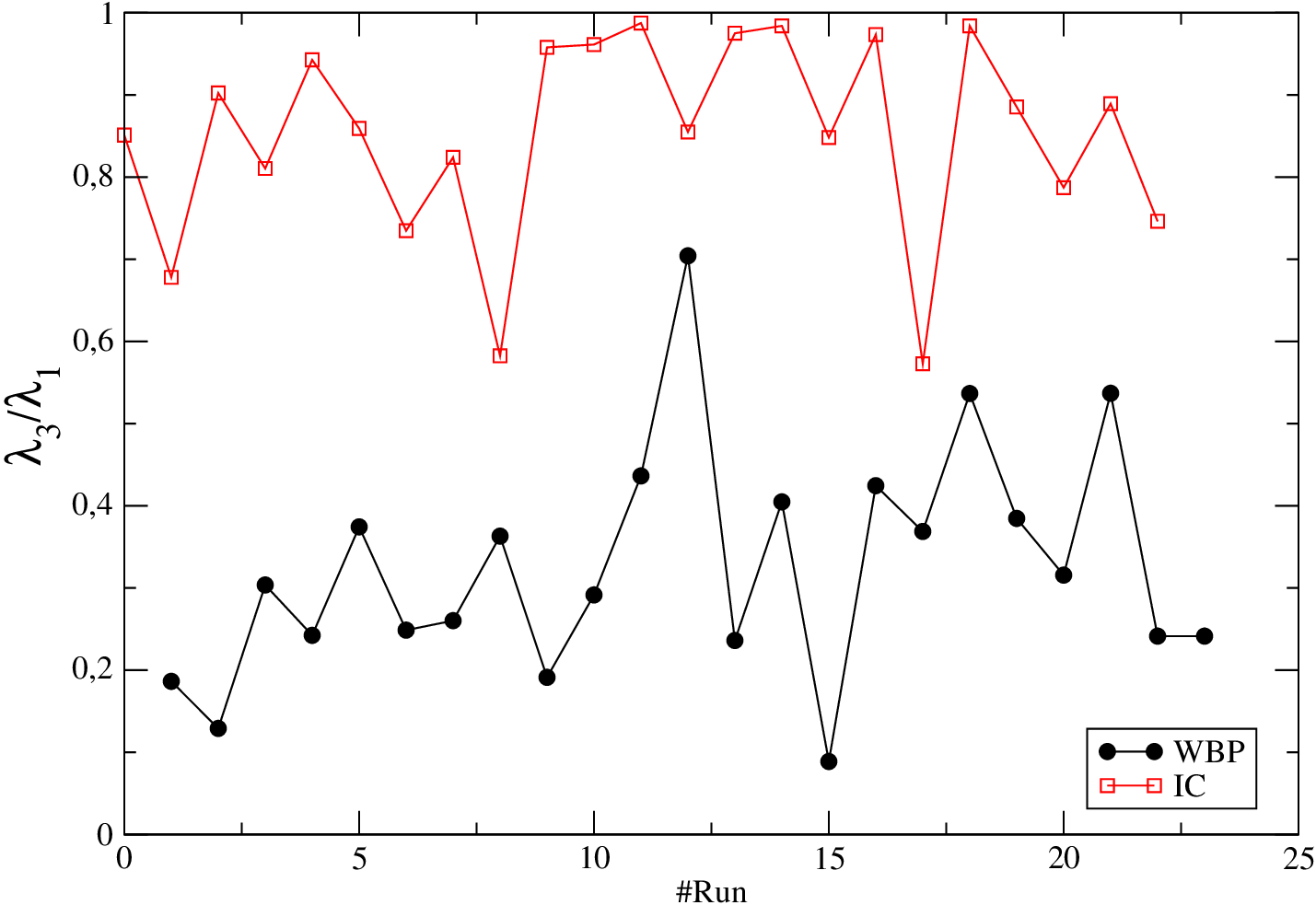}} 
\par\centering }
\caption{Ratio between the minor to major eigenvalue of the weakly
  bound particles (WBP) at time $> \tau_c$ and of the initial cloud
  (IC) in the different runs.  }
% C1
\label{c_over_a}
\end{figure}
As one may see from
Fig.\ref{c_over_a} while the initial condition was close to spherical
in all cases, the distribution of weakly bounded particles {after the collapse} 
{ becomes asymmetrical}.
Clearly satellites, being a sub-sample of the weakly bound particles,
are part of the {flattened} distribution.

{As mentioned above, the collapse of an ellipsoidal cloud is
characterized by a mechanism of energy gain that amplifies the initial
small deviations from spherical symmetry
~\citep{Benhaiem+SylosLabini_2015}.  This interpretation is supported
by the fact that the largest semi-axis (i.e. $\vec{\lambda}_3(0)$) of
the initial cloud is parallel with both that of bound particles in the
virialized state and that of ejected particles. For inhomogeneous
clouds the situation is similar and, to show that this is the case we
consider the angle $\Psi$ between $\vec{\lambda}_3(0)$ and the vector
$\vec{\lambda}^v_3(t>\tau_c)$, i.e. the smallest eigenvalue of the
virialized object at $t>\tau_c$. Furthermore we measure the angle
$\Phi$ between $\vec{\lambda}_3(0)$ and the vector
$\vec{\lambda}^{wb}_3(t>\tau_c)$ that is in the direction of the
largest semi-axis of the weakly bounded after the collapse.
Although the statistics is poor as we have only 23 runs,
Fig.\ref{pdf_psiphi} shows that the probability density function for
both $\Psi$ and $\Phi$ are both peaked at small angles. This implies
these axes are parallel among each other: this fact corroborates the
interpretation that the initial asymmetry is amplified during the VR
phase and that it leaves an imprint in the final particles
distribution.} 
\begin{figure}
\vspace{1cm} { \par\centering
\resizebox*{9cm}{8cm}{\includegraphics*{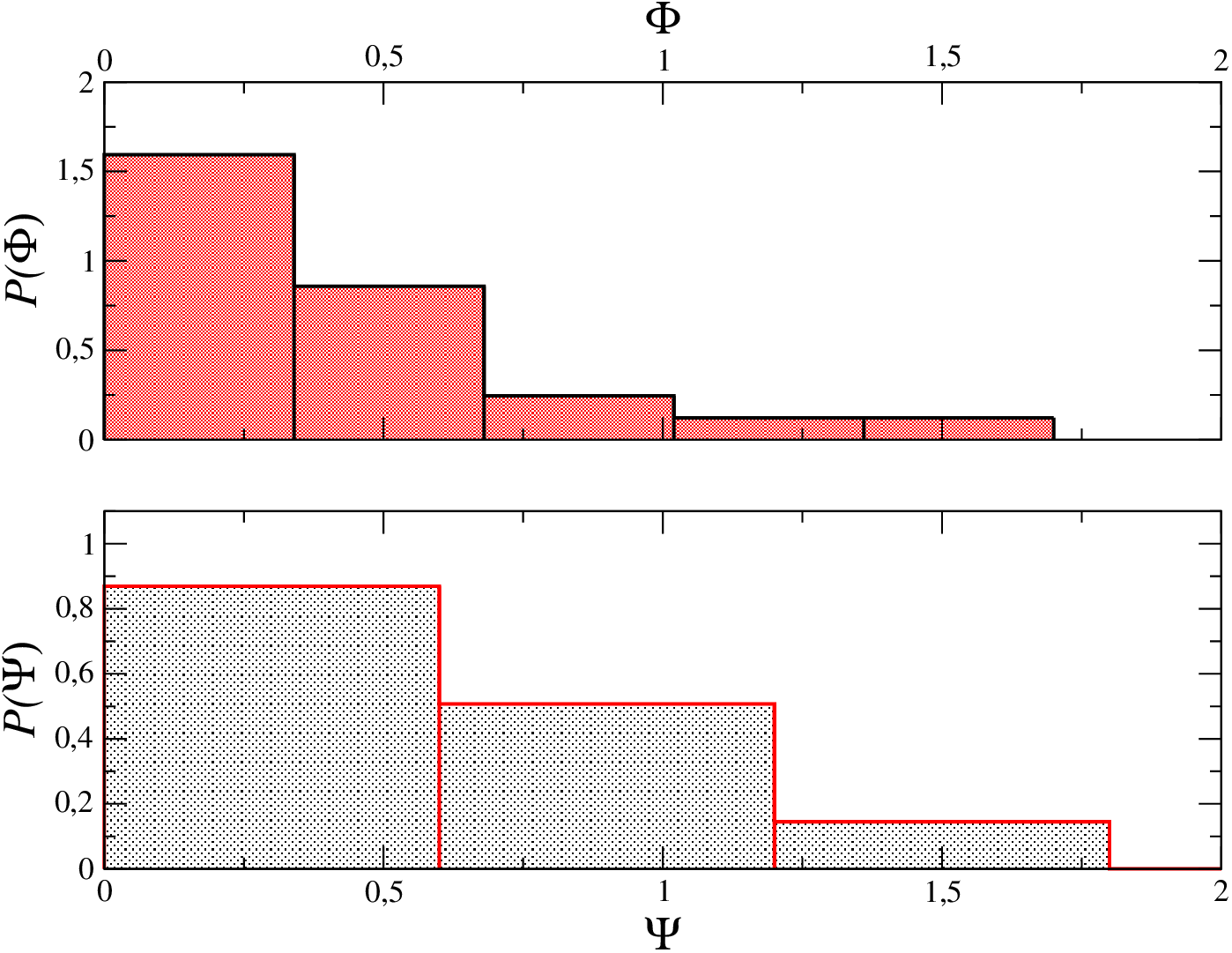}} 
\par\centering }
\caption{{Probability density function of the angles $\Psi$ and $\Phi$
  (see text for details).}}
\label{pdf_psiphi}
\end{figure}

%%%%%%%%%%%%%%%%%%%%%%%%  SATELLITE  %%%%%%%%%%%%%%%%%%%%%%%%%%%%%%%%%%%%%%%%%%%

\subsection{Satellites} 

We show in Fig.\ref{cluster_1e4_3} the evolution at different times of 
two satellites formed in a typical simulation. The satellite{s} {were}
identified at $t=7 \tau_c$ , when {they have} %it has 
approximately reached the
maximum distance from the largest virialized object, and then {their} %its
evolution was traced backward {and  forward}. One may see that
particles belonging to the satellite{s} were located, at $t=0$, in a
contiguous region in the outer shell of the initial cloud. {These}
sub-structures {are} initially not self-bound and remains bounded to
the main virialized object during {their} %its 
evolution: {their} particles have
negative, but close to zero, total energy.
During the {cold collapse phase} the sub-set of particles later
forming {these} sub-structure{s}, can gain some kinetic
energy, even though not enough to escape from the system.
{Any given satellite} reaches the largest distance allowed by its
kinetic energy and then it starts to collapse towards the main object,
which has already relaxed into a virialized quasi-equilibrium state in
a short time around $\tau_c$.  {When a certain satellite
  approaches the main virialized object, it moves in a varying
  gravitational potential and thus its total energy is not conserved.}
\begin{figure}
\vspace{1cm} { \par\centering
\resizebox*{9cm}{8cm}{\includegraphics*{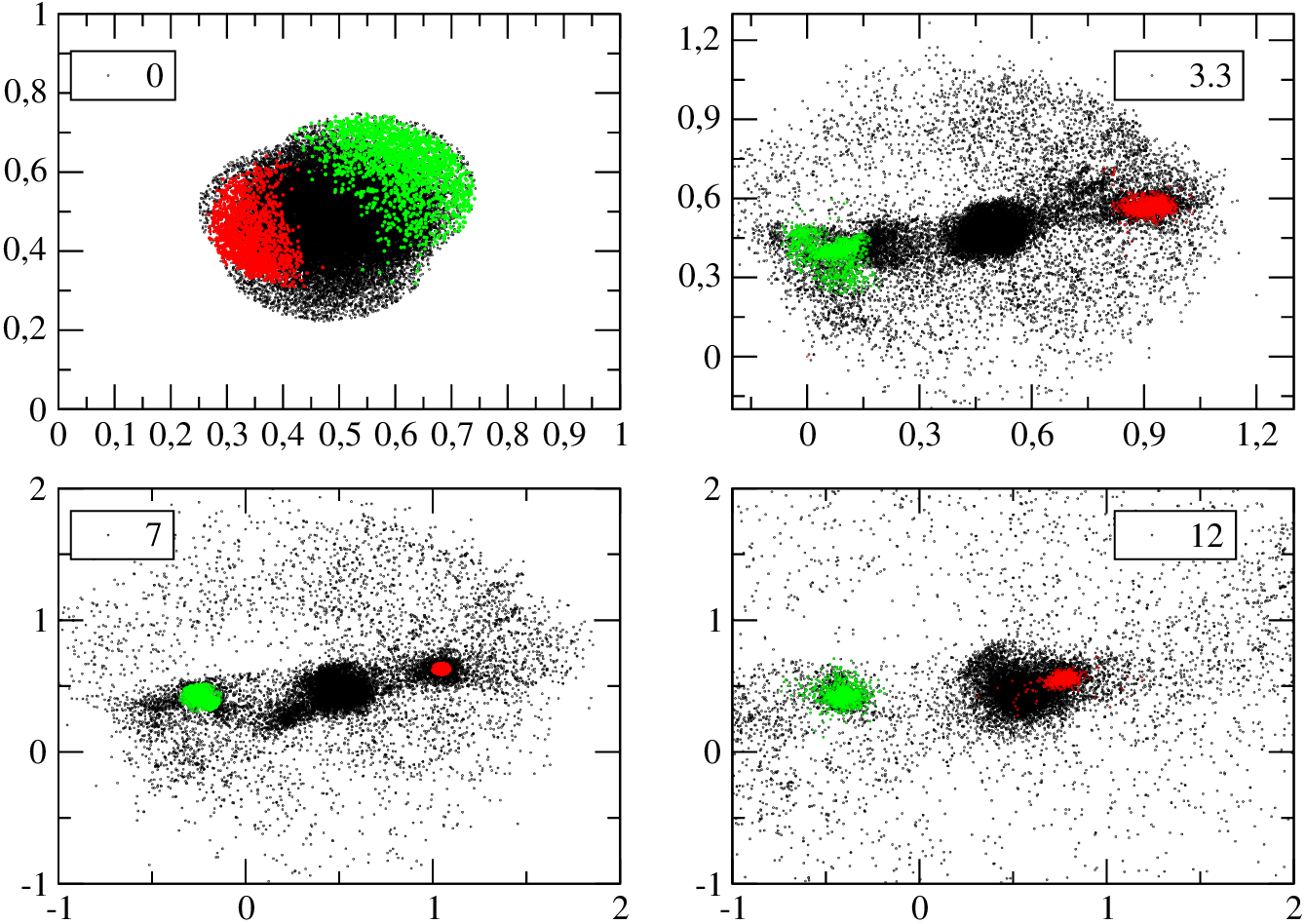}} 
\par\centering }
\caption{Evolution at different times (see labels --- times are given
  in units of $\tau_c$) of two satellites of a typical simulation (C1)
  (distances are given in units of the initial cloud radius $R_0$).
  The two satellites considered here are on the opposite side of the
  largest virialized object and have different return times. }
% C1
\label{cluster_1e4_3}
\end{figure}

Particles belonging to satellites are those which are weakly bounded
after the collapse as shown by their energy distribution 
(see Fig.\ref{cluster_1e4_12}).
\begin{figure}
\vspace{1cm} { \par\centering
\resizebox*{9cm}{8cm}{\includegraphics*{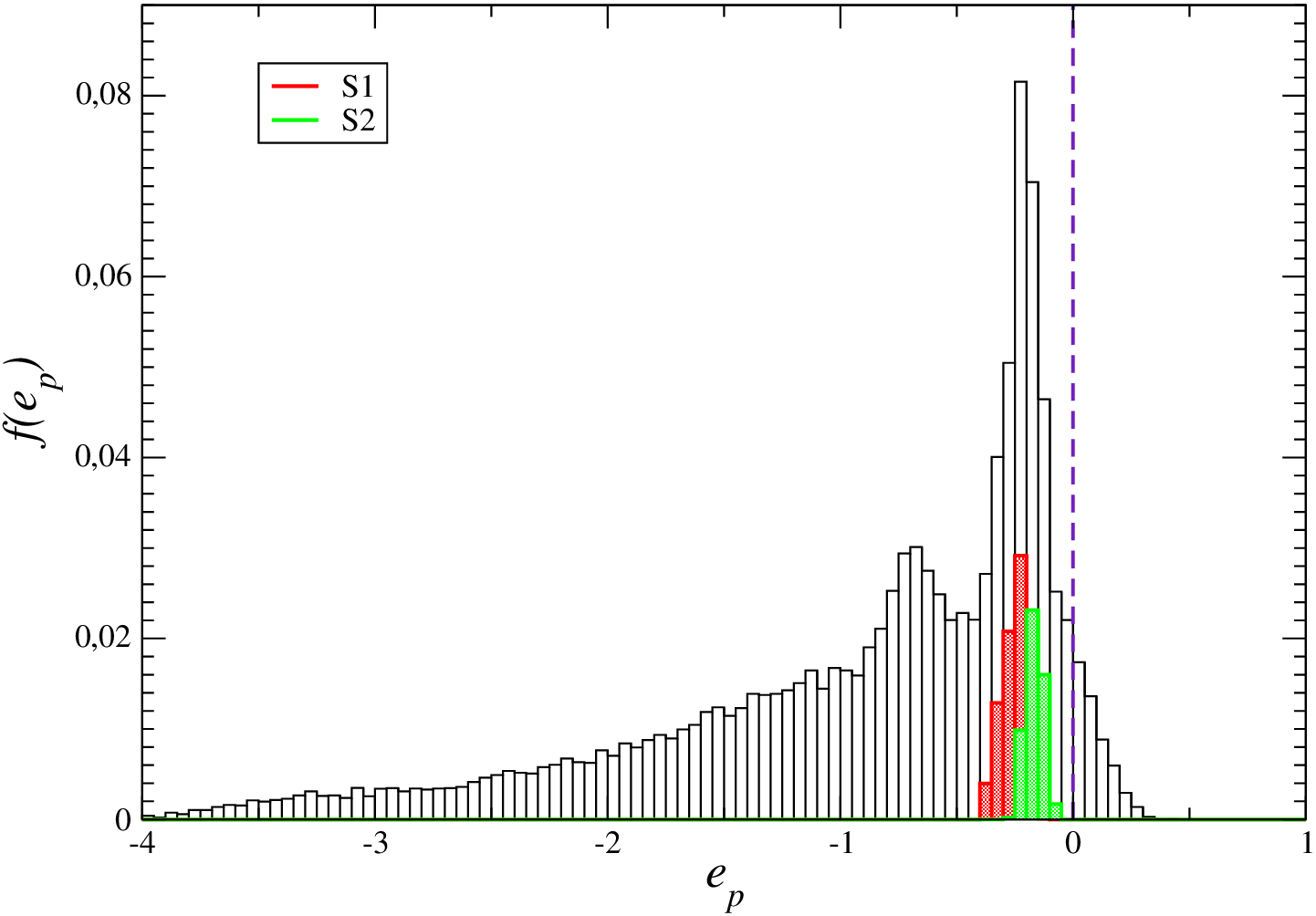}} 
\par\centering }
\caption{{Energy distribution of all particles (black) and of the
    particles belonging to the two satellites of the simulation C1
    (the color code of the satellites is the same as the one adopted
    in Fig.\ref{cluster_1e4_3}) at $t=4 \tau_c$ of a typical
    simulation.  (The energy is given in units of $e_0 = GNm^2/R_0$).
}}
\label{cluster_1e4_12}
\end{figure}
    {This situation implies that the return time $\tau_r$\footnote{
        { The time that the satellite takes to cross the main
          structure after its ejection}} of the satellite {could} be,
      {in principle,} arbitrarily long as when particle's energy is
      close to zero, i.e. $e_p \rightarrow 0^-$ then we have $\tau_r
      \rightarrow \infty$.}

The satellites' virial ratio (see the upper panel of
Fig.\ref{cluster_1e4_7}) is computed by considering the contribution
to the potential and kinetic energy of those particles identified to
belong to the satellite in the manner described above: clearly
particles' velocity is computed with respect to the satellite center
of mass rest frame.  The time $\tau_{max}$ of maximum distance from
the main object center corresponds to when the velocity of the
satellite center of mass is close to zero (see the middle panel of
Fig.\ref{cluster_1e4_7}).

{From the analysis of Fig.\ref{cluster_1e4_7}, and of the
  analogous behaviors for other satellites formed in the set of 23
  simulations, we can deduce that the satellites form at the collapse
  time of the whole cloud: this is shown by the local maximum 
  near the time of  collapse of the
  time evolution of the virial ratio and of the flatness parameter of
  the different satellites which then rapidly stabilize for a certain amount of time. 
  In addition, for the case of one satellite (shown in
  Fig.\ref{cluster_1e4_7} by the red solid line) these quantities are
  almost constant up to $\approx 12 \tau_c$ when the satellite merge
  with the main structure. Instead, the other satellite (green dashed
  line in Fig.\ref{cluster_1e4_7}), having a smaller energy (as shown
  by Fig.\ref{cluster_1e4_12}), has a return time longer than $20
  \tau_c$ and thus it is still moving away from the main structure
  when the simulation is stopped. }

\begin{figure}
\vspace{1cm} { \par\centering
  \resizebox*{9cm}{8cm}{\includegraphics*{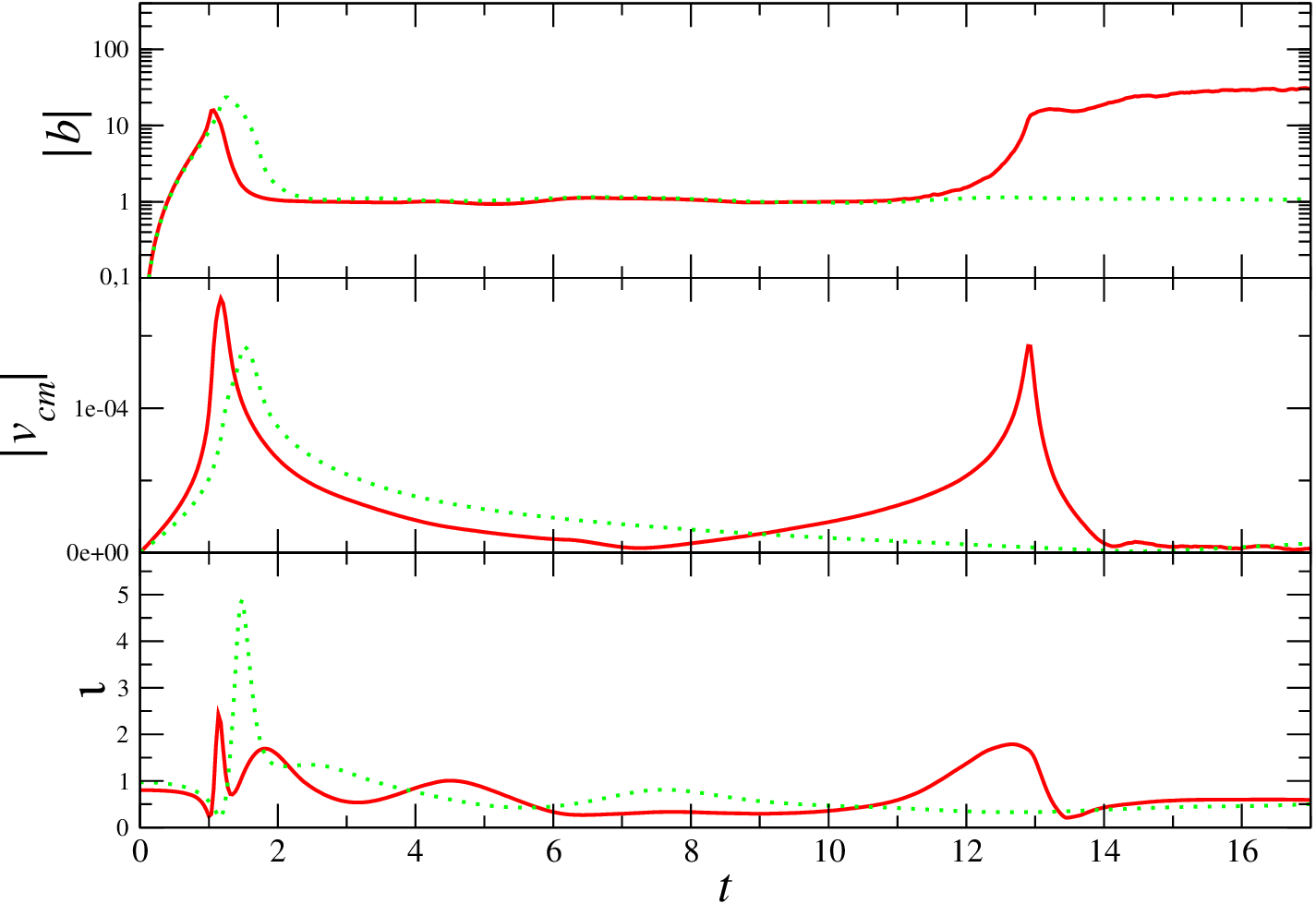}} 
\par\centering }
\caption{Upper panel: Evolution of the absolute value of the virial
  ratio for {two} different satellites of the simulation C1. Middle
  panel: velocity of their center of mass in the rest of frame of the
  main structure. Bottom panel: behavior of $\iota_{100}$, i.e.  the
  flatness parameter for all bound particles. }
\label{cluster_1e4_7}
\end{figure}

{ From the measured behaviors we note that when a satellite does
  not display a shape {roughly} close to spherical (i.e., $\iota
  \le 1$ --- see the bottom panel of Fig.\ref{cluster_1e4_7}), its
  virial ratio is far from the equilibrium value $b \approx
  -1$. Instead, a satellite may (not necessarily) reach a virialized
  configuration only when it is farther from the main object ---
  corresponding to the smallest tidal influence.  Therefore a
  satellite's particles, although being close together, may not be 
  physically bound and virially relaxed; instead they are {closely
    located} because they were so initially and they have remained
  close during the whole evolution considered.}

{A satellite might cross one or several times the largest
  virialized object before being totally destroyed by tidal
  interactions. In this situation, satellite's particles move once
  again in a rapidly varying gravitational field, and they can gain 
  energy so to escape from the system thus causing a secondary ejection 
  (see Fig.\ref{figure10}).}

\begin{figure}
\vspace{1cm} { \par\centering
  \resizebox*{9cm}{8cm}{\includegraphics*{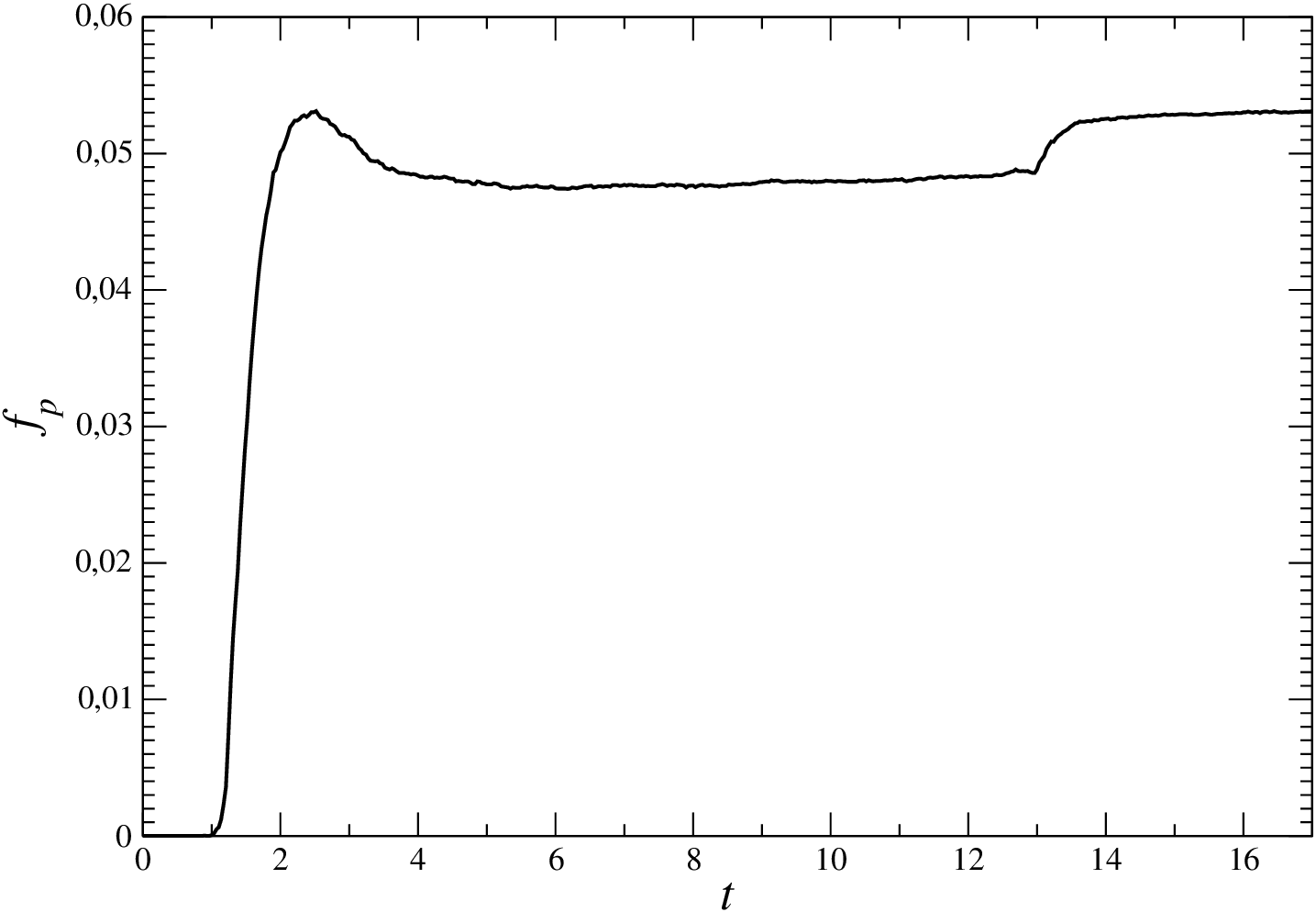}}
  \par\centering }
\caption{Fraction of { ejected particles }as a function of time for the
  simulation C1. The first secondary ejection, due to arrival of the
  first satellite to the main object, is clearly visible at $t= 13
  \tau_c$. }
\label{figure10}
\end{figure}

{ Fig.\ref{cluster_1e4_11} shows} the histogram, over all
satellites in all the simulations we have run, of $\sin(\theta)$ where
$\theta$ is the angle between the plane identified by the two largest
semi-axes of the main object and the line passing {from} the main
structure center {to} the satellite center.  This figure clearly shows
that the satellites are not only planar distributed but specifically
{they lie } in the plane {defined by the} the {two} major
axes of the main structure. This is, however, perfectly coherent with
the process we have discussed above as (1) the final major {and
  medium} axis of the main structure {are} aligned with the
initial one {and} (2) ejection occurs preferentially in the
direction of the major axis. It is therefore the position where the
probability of forming satellite is highest.

\begin{figure}
\vspace{1cm} { \par\centering
\resizebox*{9cm}{8cm}{\includegraphics*{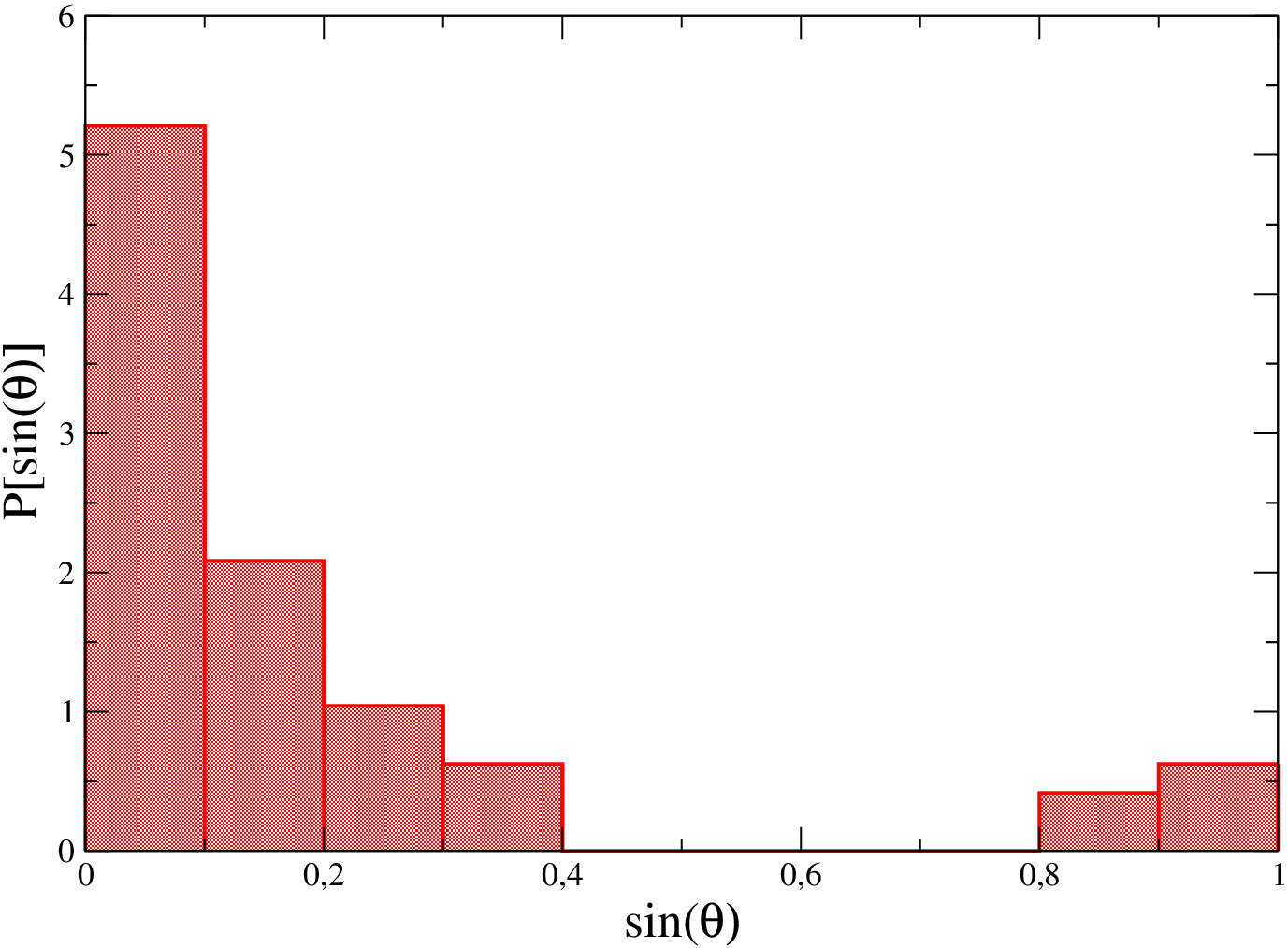}} 
\par\centering }
\caption{Probability of $\sin(\theta)$ where $\theta$ is the angle
  between the plane identified by the two largest semi-axes of the
  main structure and the line passing through the main structure
  center and the satellite center.}
\label{cluster_1e4_11}
\end{figure}

\subsection{Angular momentum} 

As mentioned above, one of the main features of the {cold collapse phenomenon} is
that a significant fraction of the system particles may gain enough
kinetic energy during the collapse phase to be ejected from it
\citep{ejection_mjbmfsl}.  In addition because the gravitational field
breaks spherical symmetry, the ejected mass can potentially carry {out} 
angular momentum.
In \cite{Benhaiem_etal_2016} for initial configurations in which
spherical symmetry {was} broken only by the Poissonian fluctuations
associated with the finite particle number $N$, we {found} that
the relaxed structures have standard “spin” parameters $\lambda
\approx 10^{-3}$ decreasing slowly with $N$. The parameter $\lambda$
is defined as \citep{Peebles_1969}
%
%\be 
%\label{lambda} 
$\lambda = {|\vec{L}_b^p|}/{L_0^b}$ 
%\ee
%
where %${L_0^b}$ is 
%
%\be
%\label{normofL2}
$L_0^b= {GM_b^{5/2} } /{\sqrt{|E_b|}} \,, $
%\ee 
%
%In Eq.\ref{normofL2}
$M_b$ is the bound mass, $E_b$ ($W_b$)
the total (gravitational potential) energy of this mass (with respect
to its center of mass) and $\vec{L}_b^p$ the angular momentum of the
bound mass.
For {prolate ellipsoidal initial conditions, with $\iota(0) \le
  0.15$,} we have measured {one order of magnitude} larger values,
i.e.  $\lambda \approx 10^{-2} $, that is of the same order of
magnitude as those reported for elliptical galaxies {(see
  \cite{Benhaiem_etal_2016} {and references therein})}.
{Naturally,} this mechanism of angular momentum generation
{can be} amplified {when} the initial distribution is already
not spherically symmetric. Indeed we found in our simulations that
typical value of the angular momentum is about $\lambda \approx
10^{-2} $ for the main virialized object (see Fig.\ref{lambda_MS}) --
although in some cases it can reach $\lambda \approx 7 \times
10^{-2}$.  {It is important to note that, while we compare with
  angular momentum measured in cosmological simulations or
  observational galaxies, we do not claim that the process we describe
  is the one taking place in these context. We compare them only in
  order to stress that the generation of angular momentum by cold
  collapse can sometimes be as effective, depending on the initial
  configuration. }
\begin{figure}
\vspace{1cm} { \par\centering
\resizebox*{9cm}{5cm}{\includegraphics*{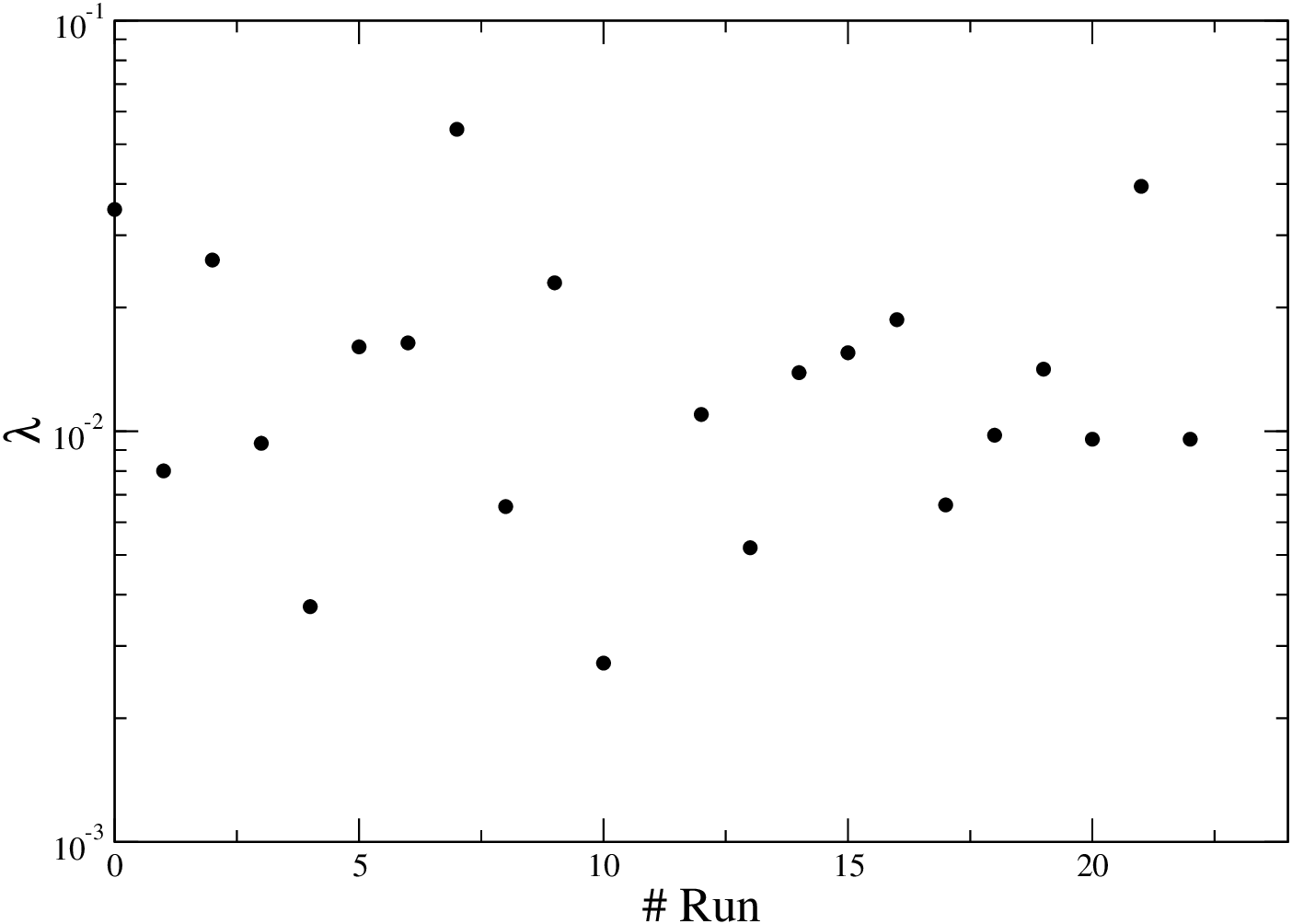}} 
%\resizebox*{9cm}{5cm}{\includegraphics*{figure12b.eps}} 
\par\centering }
\caption{Standard spin parameter ({see text}) of the main
  virialized object in 
  % (upper panel) and its satellites (bottom panel) in
  all the different runs.}
% (simulation C1)}
%Cluster 10000.1 }
\label{lambda_MS}
\end{figure}
%%%%%

%\subsection{{Summary of results}} 

%%%%%%%%%%%%%%%%%%%%%%%%%%%%%%%%%%%%%%%%%%%%%%%%%%%%%%%%%%%%%%%%%%%%%%%%%%%%%%%%%%
%%%%%%%%%%%%%%%%%%%%%%%%%%%%%%%%%%%%%%%%%%%%%%%%%%%%%%%%%%%%%%%%%%%%%%%%%%%%%%%%%%
%%%%%%%%%%%%%%%%%%%%%%%%%%%%%%%%%%%%%%%%%%%%%%%%%%%%%%%%%%%%%%%%%%%%%%%%%%%%%%%%%%

\section{Conclusions} 
\label{conclusion}

{As in the case of simpler initial conditions, such as a cold spherical
cloud, we find that, after the collapse there are two species of
particles, {identified} by their energy. However for the cases
considered here we find that, while strongly bound particles are found
in the main structure core as in the case of simpler initial
conditions previously considered, only a small fraction (i.e. less
than 20\%) of the total mass is ejected. In addition we find that
bound particles with energy $e_p \rightarrow 0^-$ may form satellite
sub-structures with a finite lifetime that can be orders of magnitude
larger than the typical time-scale $\tau_c$ of {cold collapse
  phase}.}

{These satellites are part of the {flattened} distribution made of weakly
bound particles that originates from the amplification of the initial
small deviations from spherical symmetry. Furthermore they lie
preferentially on a plane that is oriented as the plane {defined} by
the two major semi-axes of the main structure, {thus} reflecting
the initial asymmetry of the cloud. Their virial ratio is time
dependent; we observed that satellites with almost spherical shape
have virial ratio close to $b \approx -1$ when they are sufficiently
far away from the largest virialized object, while satellites with
irregular shapes have a virial ratio that can be substantially
different from the equilibrium value and are typically subjected to
large tidal effects.
}

{Our numerical experiments are greatly simplified with respect to
  cosmological simulations and, in particular, they do not consider
  any baryonic physics which is generally considered the mechanism
  that explains the formation of a disk. 
  {Nevertheless}, we note that it is
  an interesting and novel result that we have found, which is that a purely gravitational
  system may give rise to a flattened distribution of satellites.}
{It is important to stress that satellites formed during
the cold collapse phase have both a different dynamical history
than the those formed in CDM simulations and a different set up of the
initial conditions from typical cosmological initial conditions. Indeed, in the former case 
structure formation proceeds through a bottom-up, i.e. 
hierarchical, clustering process in which larger and larger objects coalesce. 
Instead, satellites produced by 
a cold collapse are formed at the same time as the main
virialized object $\tau_c$ through a monolithic collapse of the whole cloud.} 
Their lifetime depends on their energy
gain during the strong collapse phase and can be much longer than the
cold collapse typical time scale. 
Then satellites} are typically formed on opposite directions, thus
showing a {kinematic} correlation, i.e. a correlated or
anti-correlated radial velocity -- depending on whether satellites are
moving outwards or inwards. Soon after the {monolithic cold
  collapse at the time $\tau_c$} all satellites should be moving
outwards while at later times there can be satellites moving in both
directions.

{We have noticed } that satellites formed from a {cold
  collapse} are not necessarily at virial equilibrium: they can be,
{ in a certain period of their lifetime,} at virial equilibrium,
{ but in our simulation we have found that this occurs }{only}
when they are far from the main virialized object. Otherwise, even
though satellite's particles are found in a small localized spatial
region, they are not self-bound. In general a relevant fraction of the
bound particles, those which have energy close to zero, are not at
virial equilibrium. This fact can be relevant for kinematical mass
estimation and it will be studied in more detail in a forthcoming
work.

{Finally, }during the {cold collapse} a self-gravitating
system may eject a significant fraction of its mass. As we discussed
in \cite{Benhaiem_etal_2016} if the time varying gravitational field
also breaks spherical symmetry this mass can potentially carry angular
momentum. In this way starting from {initially} almost spherical
configurations with zero angular momentum can in principle lead to a
bound virialized system with non-zero angular momentum.  {We have
  shown in this paper that } %Of course this mechanism of angular
momentum generation is {even} amplified in the situation in which
the initial distribution is already not spherically symmetric.
{Further, we have noted that the collapse of the satellites into
  the main structure can also leads to an increase of the angular
  momentum's main object.}

{Despite these interesting features it is still an open problem to
  establish in which conditions a {cold collapse}  {might} occur in the
  cosmological context and/or if it does it at all.  Indeed, complex
  physical processes such as baryonic physics and merging between
  structures are not taken into account in this simple
  study. 
\bigskip 

We thank Michael Joyce and Tirawut Worrakitpoonpon for useful
discussions and comments. Numerical simulations have been run on the
Cineca Fermi cluster (project VR-EXP HP10C4S98J), and on the HPC
resources of The Institute for Scientific Computing and Simulation
financed by Region Ile de France and the project Equip@Meso (reference
ANR-10-EQPX- 29-01) overseen by the French National Research Agency
(ANR) as part of the Investissements d'Avenir program.

%%%%%%%%%%%%%%%%%%%%%%%%%

%\bibliographystyle{aa} % style aa.bst
%\bibliography{bibliography}

\end{document}